# Z-scan analysis and ab initio studies of beta-$BaTeMo_2O_9$ single crystal

I. Fuks-Janczarek[1], R. Miedzinski[1], M.G. Brik[2], A. Majchrowski[3], L.R. Jaroszewicz[3], I.V. Kityk[4]

[1]Institute of Physics, J. Dlugosz University, Al. Armii Krajowej 13/15, Czestochowa, Poland

[2] Institute of Physics, University of Tartu, Riia 142, Tartu 51014, Estonia

[3]Institute of Applied Physics, Military University of Technology, Kaliskiego 2, 00-908 Warsaw, Poland

[4]Electrical Engineering Department, Czestochowa Technological University, Al. Armii Krajowej 17/19, Czestochowa, Poland

## Abstract

The Z-scan measurements for the non-centrosymmetric optical crystals β-$BaTeMo_2O_9$ (BTMO) were performed. The corresponding experiments were carried out using the 5 ns pulses of the second harmonic of a nanosecond Nd:YAG laser at the 532 nm wavelength. It was shown that the studied crystals possess promising third-order optical susceptibilities, which allow to use the crystal as optical limiters. The comparison with other oxide materials is presented. To clarify the origin of the observed effect, the electronic and optical properties of BTMO were calculated using the density functional theory (DFT)-based method. The performed calculations of the electronic and optical properties revealed certain peculiar features that can be suitable for the non-linear optical applications. The relation between the observed nonlinear optical features and the calculated band structure is emphasized. The values of the calculated band gap and refractive index for β-$BaTeMo_2O_9$ are also reported.

**Keywords:** nonlinear optical crystals; third order nonlinear optics; band structure; ab initio calculations.

## 1. Introduction

The β-BaTeMo2O9 (BTMO) single crystals cause recently an increased attention of researchers due to interesting piezoelectric [1], polar [2], electrooptical [3], self-focusing laser generation [4] and second harmonic generation features. It is crucial for the applications that their thermal expansion possesses a weak anisotropy, although the crystals themselves have a low symmetry and the thermal conductivity of BTMO decreases with increasing temperature [5]. These crystals also were applied to a diode-side-pumped actively Q-switched intracavity Raman laser generating at 1178 nm using Nd:YAG as a gain active medium and BTMO as the Raman source [6]. Applying a pump power of 115 W and pulse repetition rate of 10 kHz, a maximum first-Stokes output power of 1.9 W was achieved. The corresponding diode-to-first-Stokes conversion efficiency is 1.7% and the first-Stokes pulse width was found to be 35 ns.

Generally most of the published articles are concentrated on the second-order optical features of crystals. However, absence of systematic studies of the third-order constants like two-photon absorption and nonlinear refractive index substantially restrain areas of applicability of non-linear optical materials. This is caused by a necessity to know how the optical limiting and self-focusing effects may influence the second order optical properties, first of all the second harmonic generation. This is particularly important when the photoinduced effects are used at different wavelengths [7]; and for the photoinduced electric polarizability [8]. As a consequence, in the present work we carry out explorations of the third-order optical features such as two-photon absorption (TPA) and nonlinear optical refractive index ($n_2$) using the Z-scan technique. We performed these studies for the 5 ns 532 nm wavelength of the second harmonic generation of the Nd:YAG laser. The measurements of the third-order optical constants may give important information about the ability to design on the base of these crystals the multi-functional devices, which allow simultaneously to transform coherently the wavelengths (first of all due to second harmonic generation (SHG)) as well as to change the light transparency, and focusing abilities. To analyse the origin of the effect bands structure calculation are performed within a framework of the Density Functional Theory (DFT).

In the second part the technology of the crystals growth is presented. Detailed description of the Z-scan method is given in Section 3. Principal results of the measurements are given in Section 4. The band structure scheme and possible relation to the Z-scan measurements are given in Section 5. After that, the paper is concluded with a short summary.

## 2. Crystal growth

$BaTeMo_2O_9$ melts incongruently at temperature equal to about 870 K. Generally the titled crystals may exist in two different structures: the high-temperature β -BTMO orthorhombic phase and the low-temperature β -BTMO monoclinic phase; they exhibit the second order optical properties, first of all for the SHG applications [9, 10]. However, to the best of our knowledge the third order optical properties of these crystals are not known and they were not yet studied.

Single crystal growth of the low-temperature β -BTMO phase was carried out by means of the Top Seeded Solution Growth (TSSG) method from self-flux high temperature solution. The starting composition of the melt was as follows: equimolar mixture of $TeO_2$ and $MoO_3$ was used as a solvent, in which 25 mol% of $BaTeMo_2O_9$ was dissolved. Owing to crystallization at low temperature gradients without pulling during the growth process, β -BTMO single crystals grew in the volume of the melt and were confined with numerous crystallographic faces. The rotation rate was 30 rpm, and the cooling rate of the melt, controlled by Eurotherm 2704 regulators, was 0.02K/h.

The optical absorption spectrum and the overall view of the studied crystals are shown in Figs. 1-2.

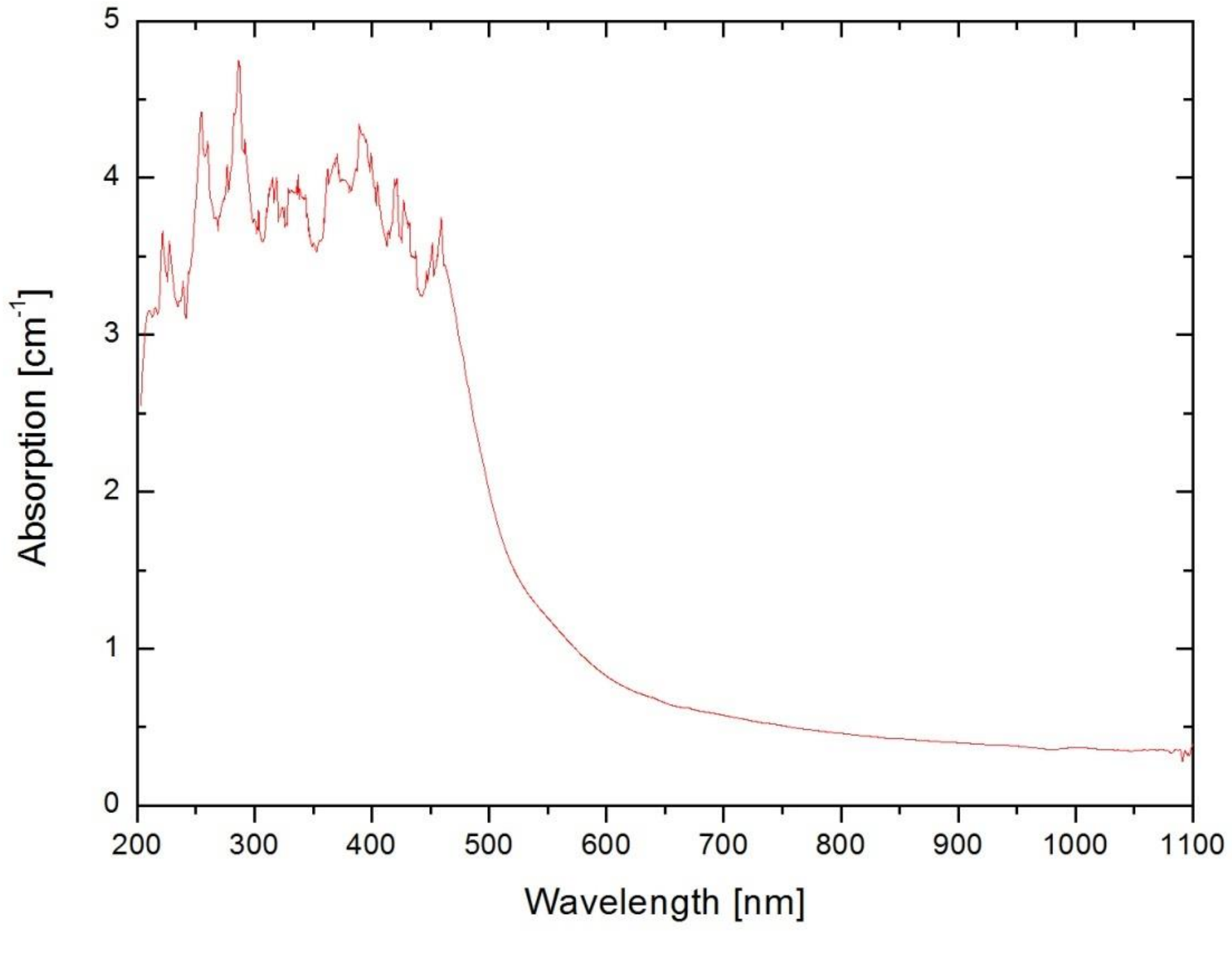

Fig. 1. Fundamental absorption edge for the β -BTMO crystals.

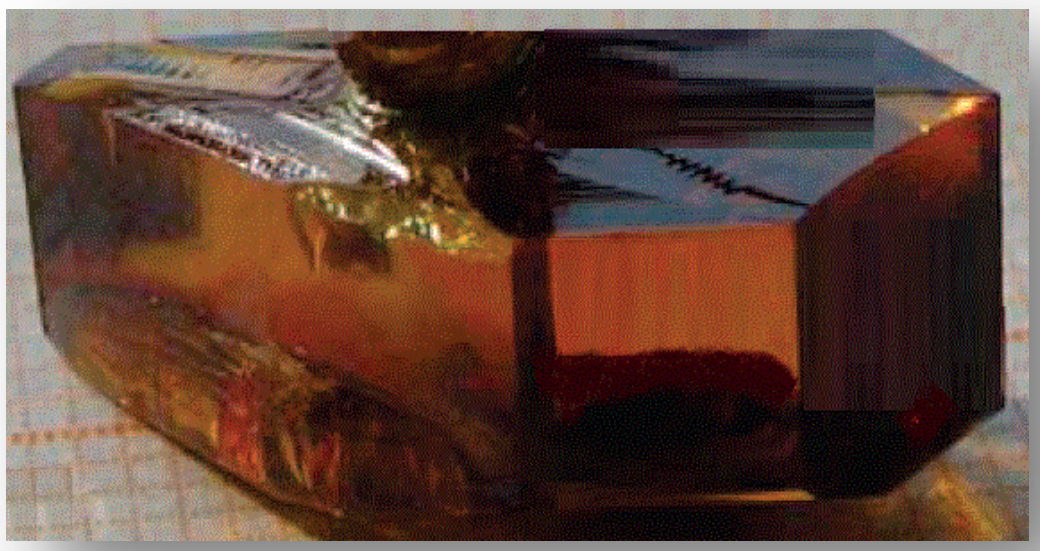

Fig. 2. Photo of the studied β –BTMO crystals.

### 3. Z-scan method

The Z-scan technique allows to do quick monitoring of the non-linear index $n_2$, nonlinearity (NLR) and the two-photon absorption (TPA) coefficient Δα in various solids [11, 12]. A Z-scan method is applied to measure both NLR and TPA in the "closed" and "open" apertures, respectively. The TPA is closely related with the multi-photon excitation and nonlinear absorption [13 – 15]. As nonlinear and multi-photon absorption can bother evaluation of the non-linear optical constants, the open aperture method is typically used together with the closed method to correct the calculated value.

In the closed aperture Z-scan configuration an aperture should restrict some undesired outside scattered light from reaching the detector. A lens focuses a laser beam to a waist point, and after this point the beam is defocused. Afterwards an aperture is placed with a detector behind it. The aperture favours such geometry that only the central region of the light cone can reach the detector. The sample is usually put at the focus point of the lens, and then is continuously moved along the $z$ axis at a distance of $z_0$ which is expressed by the Rayleigh length $z_0$:

$$z_0 = \frac{\pi \cdot W_0^2}{\lambda} \tag{1}$$

The open aperture Z-scan geometry is close to the above method; however the aperture is removed or enlarged to form the appropriate beam in the remote zone of detector. This in effect sets the normalised transmittance to $S$ = 1 to measure the non-linear absorption coefficient Δα. The main cause of non-linear absorption is the TPA processes.

If the nonlinear optical coefficient is positive, and the sample is kept behind the focus, self-focusing diminishes the signal beam divergence and thus enhances the detector signal. If the sample is moved to the left-hand side of the focus, the focus is shifted to the left, and the stronger divergence after the focus suppresses the detector signal. Following the obtained dependence of the detector signal intensity on the sample position, the magnitude of the nonlinear index is evaluated. Note that nonlinear absorption, including that one which is due to the TPA, may also disturb the measured signal. This can be measured separately by measuring the power of the whole transmitted laser beam. Taking into account these data, the measurement of nonlinearity can be corrected [11, 12].

The Z-scan presents a single beam approach evaluation of third order nonlinear refractive index $n_2$ and TPA values [11 – 13]. The relation between the induced phase distortion $\Delta\Phi_0$ and $\Delta T_{V\text{-}P}$ (difference between the normalized valley and maximum transmittances) for the third-order NLO refractive process is:

$$\Delta T_{V-P} \cong 0.406(1-S)^{0.27}|\Delta\Phi_0| \quad (2)$$

where

$$\Delta\Phi_0 = \frac{2\pi}{\lambda} \mathrm{n}_2 I_0 (1-\mathrm{e}^{-\alpha L})/\alpha \quad (3)$$

and $S$ is the aperture transmittance without sample, $n_2$ – the third order nonlinear refractive index, $I_0$ – peak on axis irradiance at the focus, α - absorption coefficient, λ – wavelength, $L$ – the sample length.

The transparency of the thick nonlinear sample (a larger sample length $L$) may be described by the following expression [11]:

$$T(X) \approx 1 + \Delta\phi_{Z_0} F(x,l) \quad (4)$$

where

$$F(x,l) = \frac{1}{4} ln\left(\frac{([x+l/2]^2+1)([x-l/2]^2+9)}{([x-l/2]^2+1)([x+l/2]^2+9)}\right) \quad (5)$$

Here the $\Delta\phi_{Z_0} = (2\pi/\lambda) n_2 I_0 Z_0$.

A crucial role here is played by the diffraction length, $Z_0$, which for the focused beam is determined as $\pi w_0^2/\lambda$ for a Gaussian beam with $w_0$ being the waist diameter. The position of the aperture is not critical for a large distance from the focus, $d \gg Z_0$. The typical values of these parameters are varied from $20Z_0$ to $100Z_0$. The parameter of the renormalized length may be introduced as $x = Z/Z_0$ and the sample's geometry factor $l = L/Z_0$, where $Z$ is a distance between the sample and the focal point. For the open aperture where S = 1 the TPA coefficient α can be quickly estimated by fitting the following expression

$$\Delta T(Z) \approx -\frac{q_0}{2\sqrt{2}}\frac{1}{\left(1+\frac{Z^2}{Z_0^2}\right)} \quad (6)$$

where

$$q_0 = \beta I_0(1 - \mathrm{e}^{-\alpha L})/\alpha \quad (7)$$

to the experimental data points.

The experimental setup is shown in Fig. 3. The continuum Minilite II Laser was used as a source of the Gaussian beam.

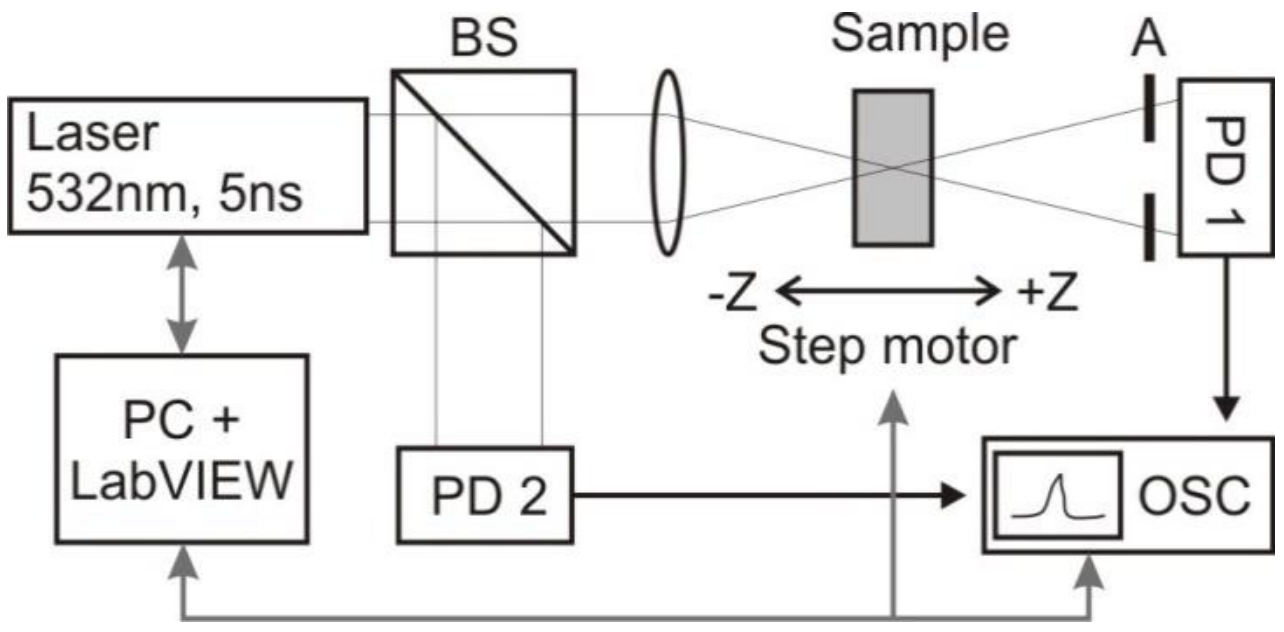

Fig. 3. Z-Scan experiment setup. BS- beam splitter, A – aperture, PD1, PD2 – photodetectors, OSC – Oscilloscope.

The sample was moved along the axis of the focused laser beam with the increment of 1 μm. The experimental set-up was operated by LabVIEW program and all the data were stored on the computer.

**4. Experimental results and discussion**

Fig. 4 shows an closed aperture Z-Scan curve of the 7 mm thick BTMO crystal, which was obtained using the 5 ns pulse of a Nd:YAG laser having an energy of 0.75 mJ at 532 nm wavelength. The Gaussian beam spot radius at waist focus was $w_0$ = 18 μm, which corresponded to power density equal to $I_0$ = 4.6 GW/cm$^2$. The crystal has a positive nonlinearity (self-focusing), because the curve configuration is a valley – peak having a change in the normalized transmittance $\Delta T_{V\text{-}P}$ = 0.68. According to this result, the measured nonlinear refractive index change is equal to about $n_2 \approx 5.5 \times 10^{-5}$. Under the same condition, the Z-scan with 100% open aperture, as shown in the figure, exhibits the TPA spectrum of the examined crystal. Following the performed theoretical fit (solid line) the evaluated value of the TPA β= 0.34 cm/GW for wavelength 532 nm.

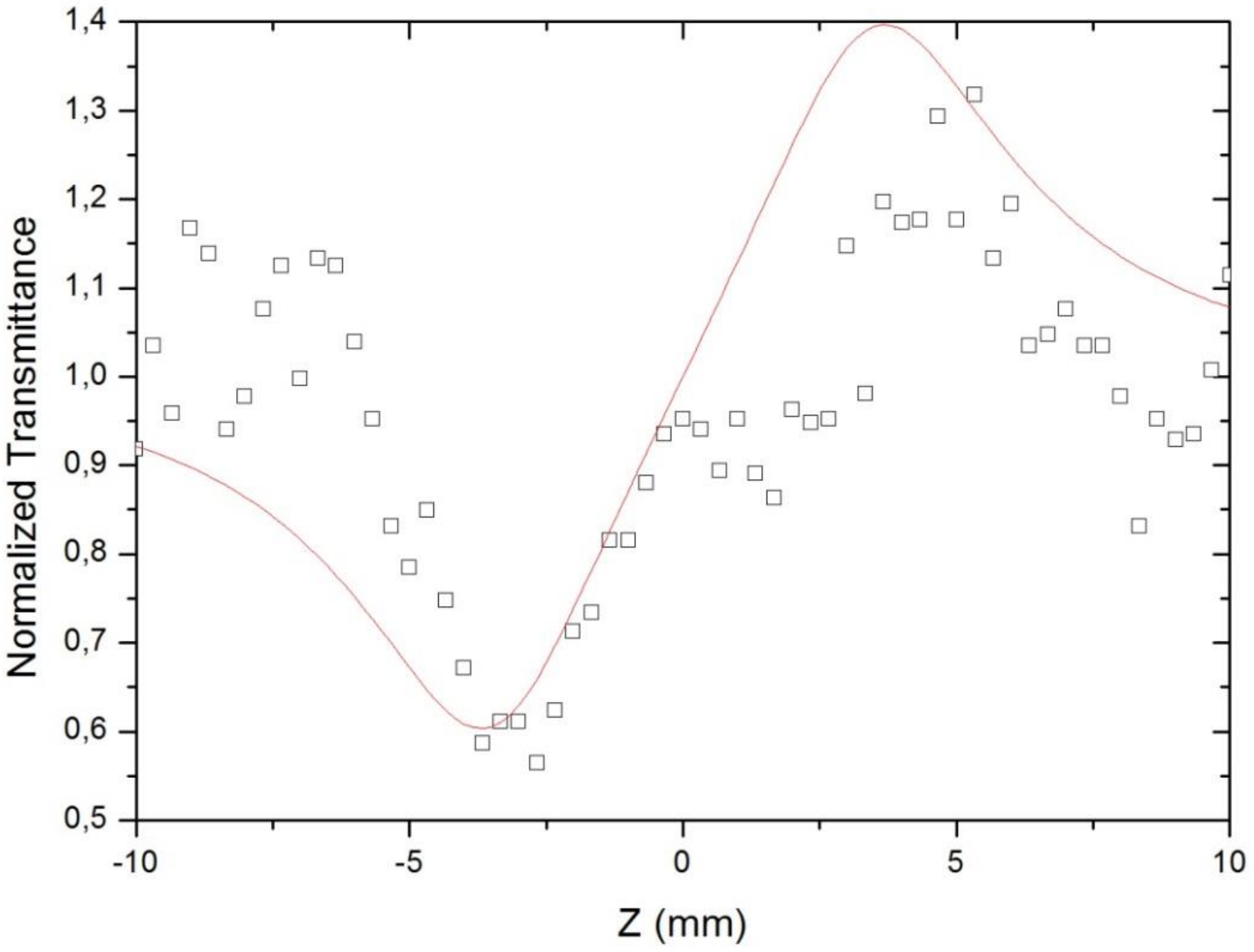

Fig. 4. Measured close aperture Z-scan for the β-BTMO crystal.

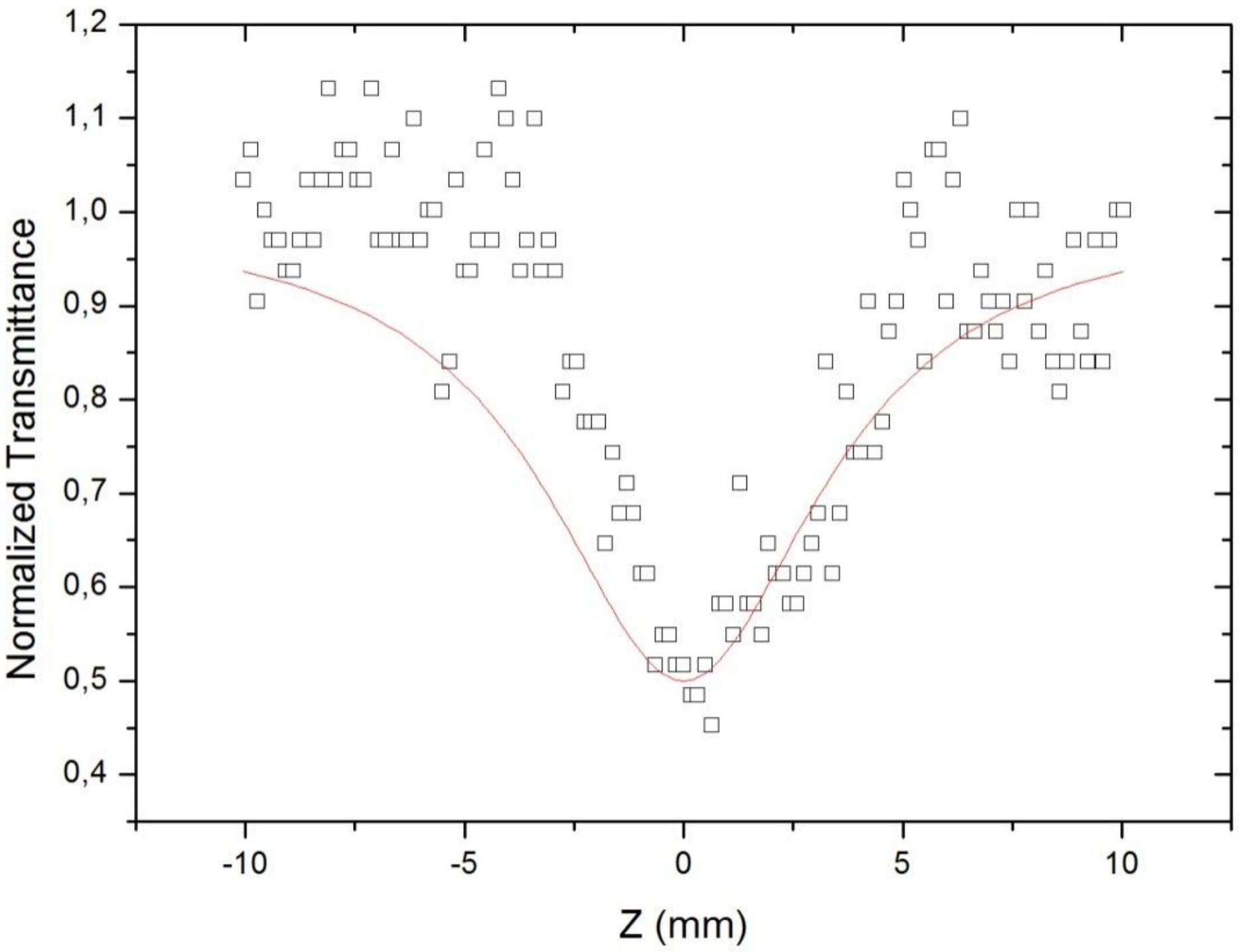

Fig. 5. Measured open aperture Z-scan of the β–BTMO crystal. Squares – experimental data, solid line – the theoretical result with data fit using β = 0.34 cm/GW.

The performed measurements have given the magnitudes of β equal to 0.34 cm/GW and $n_2 = 1.19\times10^{-14}$ $cm^2/W$ at the 532 nm wavelength. With respect to other molybdates, this value is a moderate one. This fact is very important for the applications. For example, for $PbMoO_4$ this value is equal to 2.4 cm/GW, however for the $CaMoO_4$ this value is equal to 0.14 cm/GW at 532 nm [16]. To clarify such behaviour of the studied BTMO crystal, in the next section we report the results of the performed band structure calculations, which allowed for finding some relation between the experimentally found non-linear properties and band structure parameters.

## 5. Calculations of the electronic band structure and origin of nonlinear optical features

The CASTEP module [17] of Materials Studio package was used to calculate the structural, electronic and optical properties of $BaTeMo_2O_9$. The initial structural data for this compound were taken from Ref. [18]. The generalized gradient approximation (GGA) with the Perdew-Burke-Ernzerhof [19] and the local density approximation (LDA) with the Ceperley-Alder-Perdew-Zunger (CA-PZ) functional [20, 21] were used to treat the exchange-correlation effects. The Monkhorst-Pack k-points grid was chosen as 3×2×2. The cut-off energy, which determines the size of the plane-wave basis set, was set at 340 eV. The convergence criteria were as follows: $10^{-5}$ eV/atom for energy, 0.03 eV/Å for maximal force, 0.05 GPa for maximal stress and $10^{-3}$ Å for maximal displacement. The electronic configurations were as follows: $5s^25p^66s^2$ for Ba, $5s^25p^4$ for Te, $4s^24p^64d^55s^1$ for Mo, and $2s^22p^4$ for O, and the ultrasoft pseudopotentials were used for all chemical elements.

Table 1 collects a summary of the experimental and calculated structural properties of the β-BTMO crystal. As seen from the Table, agreement between the calculated and experimental lattice parameters is good. As usual, the GGA-obtained results somewhat overestimate the experimental lattice parameters, whereas the LDA-calculated values are a little underestimated if compared to the experimental data.

Table 1. Summary of calculated structural constants and band gap for the β-BTMO crystal

| | Experim. [18] | Calc. (this work) | | Calc. [22] |
|---|---|---|---|---|
| | | GGA | LDA | |
| *a*, Å | 5.53460 | 5.73606 | 5.38114 | 5.6849 |
| *b*, Å | 7.45620 | 7.62457 | 7.40990 | 7.5721 |
| *c*, Å | 8.83420 | 8.98262 | 8.69990 | 8.9373 |
| β, ° | 90.8970 | 90.2447 | 90.2549 | - |
| *V*, $Å^3$ | 364.51 | 392.851 | 346.894 | 348.71 |
| Band gap, eV | - | 2.669 | 2.610 | 2.78 |

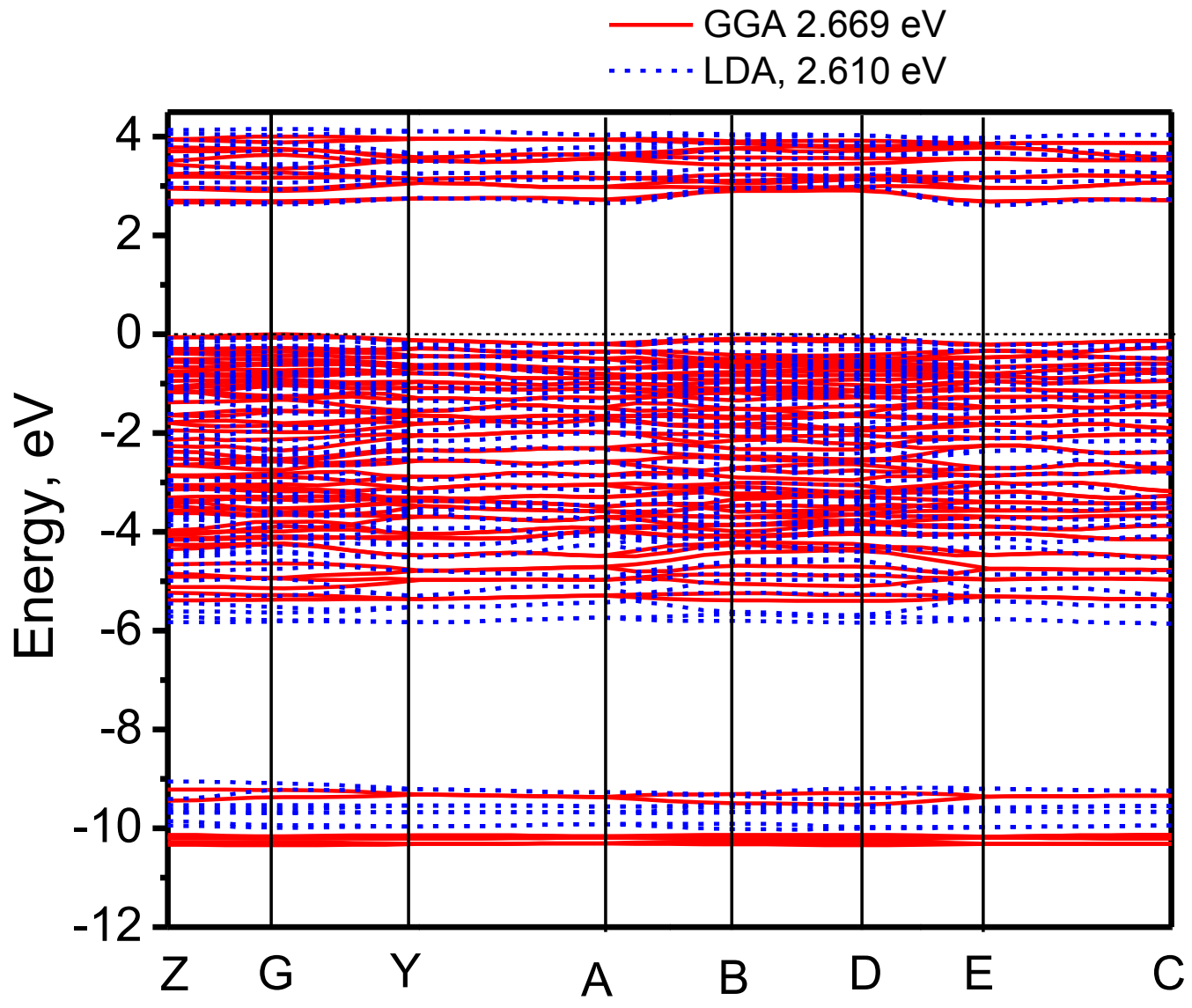

Fig. 6. Calculated band energy structure of the β-BTMO crystals. The calculated band gaps are also given in the figure.

The calculated band structure of the β-BTMO crystal is shown in Fig. 6. Following Fig. 6 one can see that the most of the electronic bands possess the flat dispersion in the k-space. The only exception presents the D-E-C direction in the Brillouin zone (BZ). Existence of such segment in the BZ may play a principal role for mobility of the electrons in the conduction band and holes in the valence band.

Generally the band structure defines the values of the TPA coefficient through the existence of the double resonances; at fundamental and doubled frequency wavelengths [23, 24]. The effect will be moderated, because there will be some resonance between the trapping levels, the fundamental band gap and the two-photon states in the conduction band. The later should have the different values of the resonances for different points of the BZ. Additionally the phonon subsystem may produce the anharmonic electron-phonon contribution, providing in this way an additional basis for the sufficient TPA signal far from the resonances.

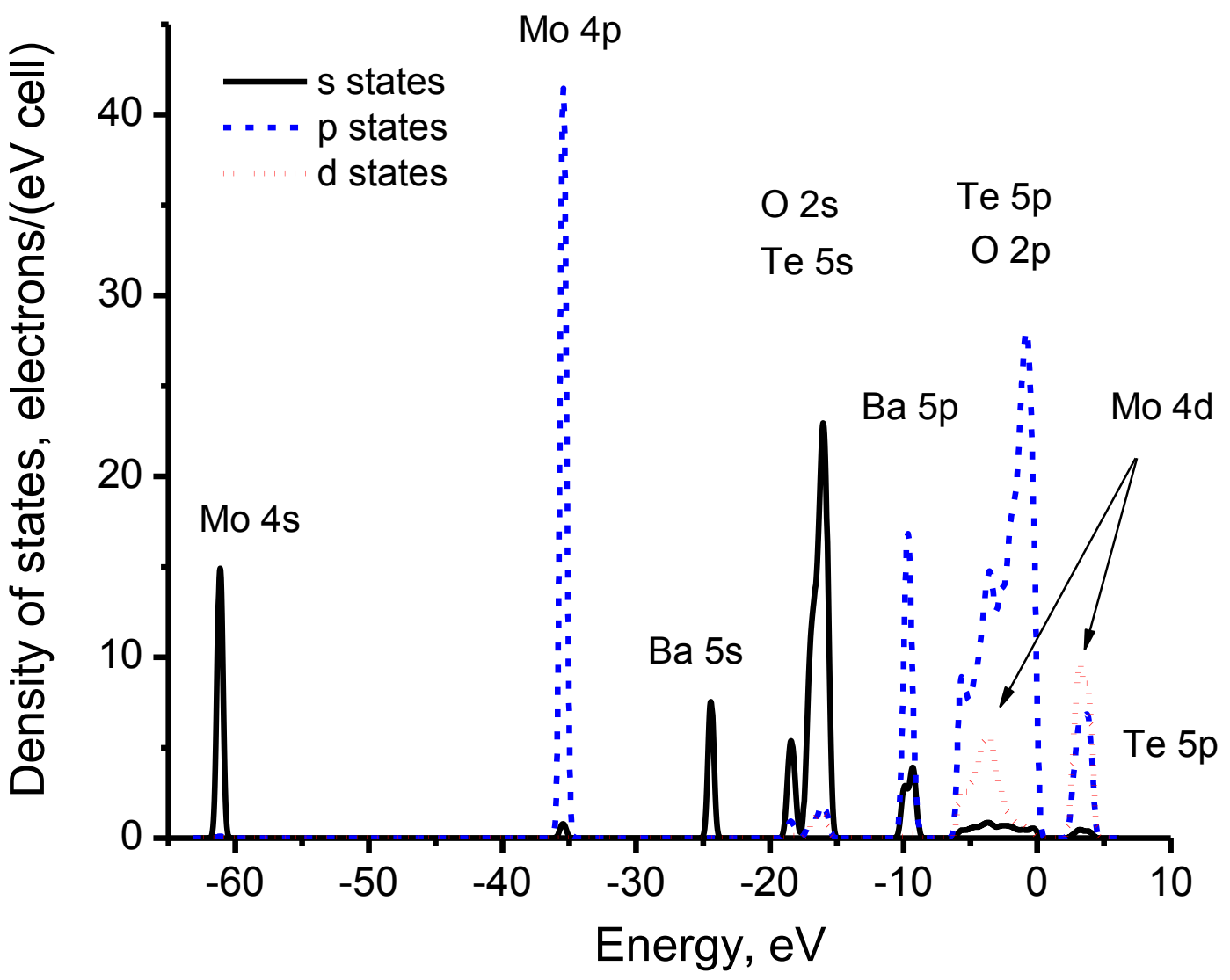

Fig. 7. Partial density of states for the studied β-BTMO single crystals.

Composition of the calculated band structure from Fig. 6 can be understood by looking at the density of states (DOS) diagrams, presented in Fig. 7. A narrow (about 1.5 eV) conduction band is composed mainly of the Te 5p and Mo 4d states. The valence band, whose width is about 5 eV, is dominated by the oxygen 2p states, with a small admixture of the Mo 4d states due to the hybridization effects. A narrow band with a sharp peak at about -10 eV is made by the Ba 5p states, and doubly-peaked band between -19 eV and -15 eV is formed by the O 2s and Te 5s states. There are three very deep in energy bands: the Ba 5s states (-15 eV), the Mo 4p states (-36 eV) and the Mo 4s states (-61 eV); they do not manifest themselves in the optical spectra.

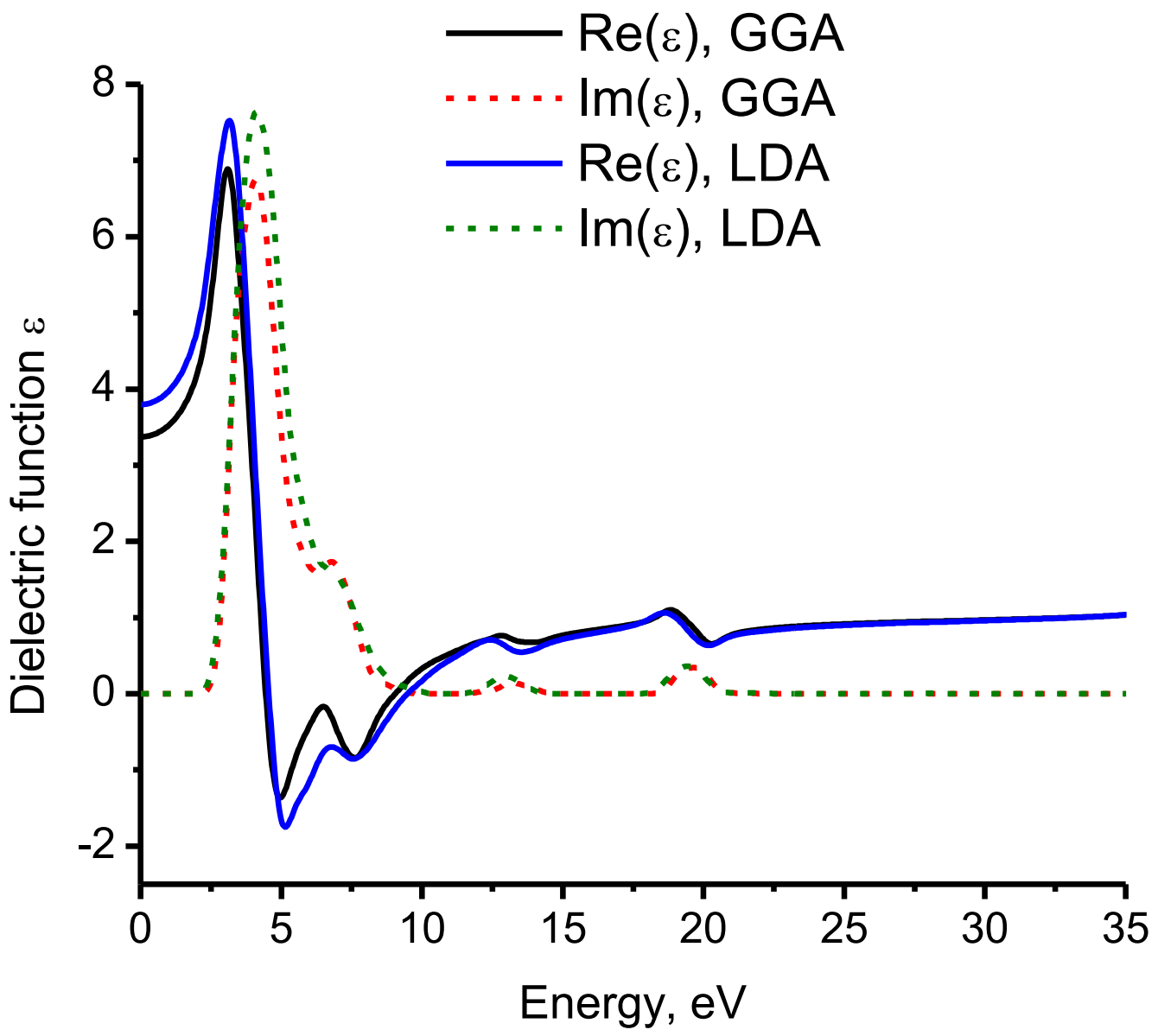

Fig. 8. Dispersion of the real Re(ε) and imaginary Im(ε) dielectric function for the β-BTMO crystals.

Fig. 8 shows the calculated dielectric function of β-BTMO crystals. The imaginary part Im(ε(ω)) of a dielectric function ε(ω) (directly related to the absorption spectrum of a solid) was calculated by numerical integrations of dipole matrix operator elements between all possible combinations of the occupied states in the valence band and empty states in the conduction band:

$$\mathrm{Im}(\varepsilon(\omega)) = \frac{2e^2\pi}{\omega\varepsilon_0}\sum_{k,v,c}\left|\left\langle \Psi_k^c \middle| \vec{u}\cdot\vec{r} \middle| \Psi_k^c \right\rangle\right|^2 \delta\left(E_k^c - E_k^v - E\right), \tag{8}$$

where $\vec{u}$ is the polarization vector of the incident electromagnetic field; $\vec{r}$ and $e$ are the electron's position vector and electric charge, respectively, $\Psi_k^c$, $\Psi_k^c$ are the wave functions of the conduction and valence bands at BZ k point, respectively; $E = \hbar\omega$ is the incident photon's energy; $\varepsilon_0$ is the vacuum dielectric permittivity. The summation in Eq. (8) is carried out over all states from the occupied and empty bands, with their wave functions obtained after the crystal structure has been optimised to achieve proper convergence of the calculated results.

The real part Re(ε(ω)) of the dielectric function ε, which determines the dispersion properties and refractive index values, can be estimated then by using the Kramers-Kronig relation [25]:

$$\mathrm{Re}(\varepsilon(\omega)) = 1 + \frac{2}{\pi}\int_0^\infty \frac{\mathrm{Im}(\varepsilon(\omega'))\omega' d\omega'}{\omega'^2 - \omega^2} \tag{9}$$

From the analysis of the optical function (Fig. 8), one can conclude that the main maximum of the imaginary part of the dielectric function lies at energies equal to about 2.8 eV. The latter are not far from the excitation of the two-photon states. At the same time one can observe the changes of the signs for the real part for the dielectric susceptibility, which, as a rule, is caused by occurrence of the plasmon excitations. The latter leads to variation of the free carriers and corresponding screening effects [26], which define the values of the ground state dipole moments directly related to the third order hyperpolarizabilites. The behaviour is principally different from many other oxide band structures and corresponding optical functions [27 – 29] and following Fig. 7, the strongly localized Te 5p, 4p and Mo 4d states are effectively hybridized with the delocalised anionic p-states. So varying appropriately the number of the intrinsic defects one can achieve the desired value for the TPA and the nonlinear refractive indices which define the optical limiting and beam self-focusing features.

As an additional result, which can be extracted from Fig. 8, we report the calculated values of the refraction index for the β-BTMO crystals as $n$=1.836 (GGA) and $n$= 1.949 (LDA), estimated as a square root of Re(ε) in the limit of the infinite wavelength (zero energy).

## 6. Conclusions

For the first time the Z-scan measurements of the acentric β-BTMO single crystals were performed. The measured nonlinear refractive index change was found to be equal to about $n_2 \approx 5.5\times10^{-5}$. The evaluated value of the TPA β= 0.34cm/GW for the 532 nm wavelength, which is optimal for possible application in different optoelectronic devices and seems to be advantageous if compared to other molybdates. The origin of the observed effect is explained by specific features of the band structure dispersion. There are some resonances between the trapping levels in the fundamental band gap and the two-photon states in the conduction band. Additionally there exists a substantial anisotropy of the carrier mobility, and the occurrence of the plasmon resonances giving additional number of carriers is near the two-photon resonances. The origin of the effect is caused by strongly localized Te 4p, 5p and Mo 4d states, which effectively are hybridized with the delocalised anionic p-states.

The performed first principles calculations of the electronic band structure have shown the β-BTMO single crystals to have a direct band gap, whose numerical value was 2.669 eV (GGA) and 2.610 eV (LDA). Since the GGA/LDA calculated band gaps are, as a rule, always underestimated, the above-given numbers represent the lowest estimate of the band gap, qhose real value can be about 3.6-4 eV. In addition, from the calculated optical properties of the β-BTMO crystals we could estimate the index of refraction *n* as 1.836 (GGA) and 1.949 (LDA).

The electronic and non-linear properties of the studied BTMO crystals are principally different from many other oxide band structures and corresponding optical functions and this is a consequence of the strongly localized Te 5p, 4p and Mo 4d states which are effectively hybridized with the delocalised anionic p-states. So varying appropriately the number of the intrinsic defects one can achieve the desired value for the TPA and the nonlinear refractive indices, which define the optical limiting and beam self-focusing features

## Acknowledgments

We acknowledge support of the Polish Ministry of Sciences and Higher Education, Key Project POIG.01.03.01-14-016/08-08 "New Photonic Materials and Their Advanced Application" in 2013.M. G. Brik acknowledges support from i) European Social Fund's Doctoral Studies and Internationalization Program DoRa and ii) European Union through the European Regional Development Fund (Center of Excellence ''Mesosystems: Theory and Applications'', TK114). Dr. G.A. Kumar (University of Texas at San Antonio) is thanked for allowing to use Materials Studio.